\documentclass[11pt,letterpaper]{article}
\usepackage{graphicx}
\usepackage{subfigure}
\usepackage{epsfig}
\usepackage{amsmath}

\newcommand{\BABARPubYear}    {06}

\newcommand{\BABARConfNumber} {026}
\newcommand{\SLACPubNumber} {12000}

\input dir_B2LLG/babarsym
 
\long\def\inst#1{\par\nobreak\kern 4pt\nobreak
    {\it #1}\par\vskip 10pt plus 3pt minus 3pt}

\begin{document}
{\pagestyle{empty}

\begin{flushright}
\babar-CONF-\BABARPubYear/\BABARConfNumber \\
SLAC-PUB-\SLACPubNumber \\
\end{flushright}

\par\vskip 5cm

\begin{center}
\Large \bf A Search for the \boldmath \beeg and \bmmg Decays
\end{center}
\bigskip

\begin{center}
\large The \babar\ Collaboration\\
\mbox{ }\\
\today
\end{center}
\bigskip \bigskip

\begin{center}
\large \bf Abstract
\end{center}
With the \babar detector at the PEP-II asymmetric \B Factory at SLAC, we 
present the first search for the decays \bllg ($\ell=e$, $\mu$). Using 
a data set of \lumi\invfb collected at the \FourS resonance, we find no 
significant signal and set the following branching fraction upper limits 
at 90\% confidence level: $\BR(\beeg)< \ResultE\times 10^{-7}$ and 
$\BR(\bmmg)< \ResultM\times 10^{-7}$.

\vfill
\begin{center}

Submitted to the 33$^{\rm rd}$ International Conference on High-Energy Physics, ICHEP 06,\\
26 July---2 August 2006, Moscow, Russia.

\end{center}

\vspace{1.0cm}
\begin{center}
{\em Stanford Linear Accelerator Center, Stanford University, 
Stanford, CA 94309} \\ \vspace{0.1cm}\hrule\vspace{0.1cm}
Work supported in part by Department of Energy contract DE-AC03-76SF00515.
\end{center}

\newpage
} 

\begin{center}
\small

The \babar\ Collaboration,
\bigskip

%% author list as of 01-Jul-2006 (596 authors)
%
{B.~Aubert,}
{R.~Barate,}
{M.~Bona,}
{D.~Boutigny,}
{F.~Couderc,}
{Y.~Karyotakis,}
{J.~P.~Lees,}
{V.~Poireau,}
{V.~Tisserand,}
{A.~Zghiche}
\inst{Laboratoire de Physique des Particules, IN2P3/CNRS et Universit\'e de Savoie,
 F-74941 Annecy-Le-Vieux, France }
% ...

\end{center}\newpage

\section{INTRODUCTION}
\label{sec:Introduction}

Rare decays induced by flavor changing neutral currents occur at loop level in the Standard Model (SM) and they are sensitive to the 
flavor structure of the SM as well as to the new physics beyond the SM. Studying radiative \B meson decays such as \bllg
\footnote{Throughout the document, $\ell$ denotes either $e$ or $\mu$, and the charge conjugated states are implicitly included.} 
can provide us essential information on the parameters of the SM, such as the elements of the Cabibbo-Kobayashi-Maskawa matrix 
and the leptonic decay constants. As explained below, the branching ratios that the SM predicts are much below the experimental 
sensitivity. Hence this paper presents a search for new physics.

\begin{figure}[!htb]
\begin{center}
\subfigure[]{\epsfig{file=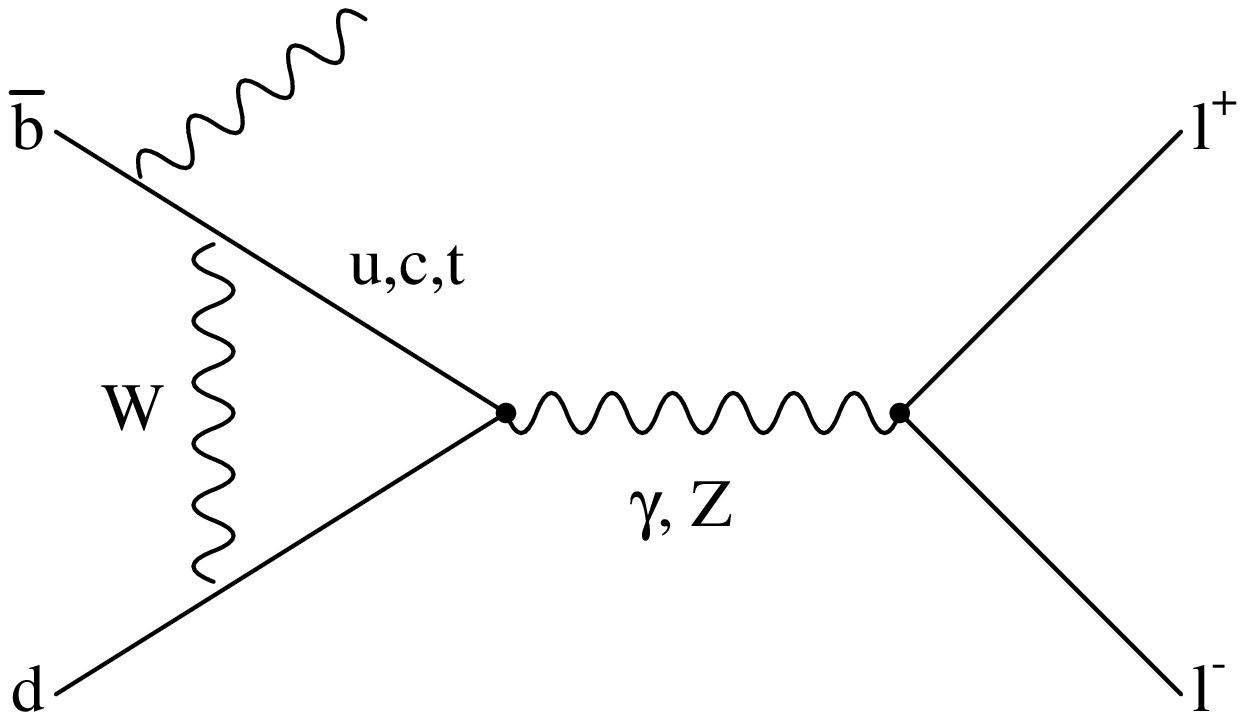, width=0.3\textwidth}}
\subfigure[]{\epsfig{file=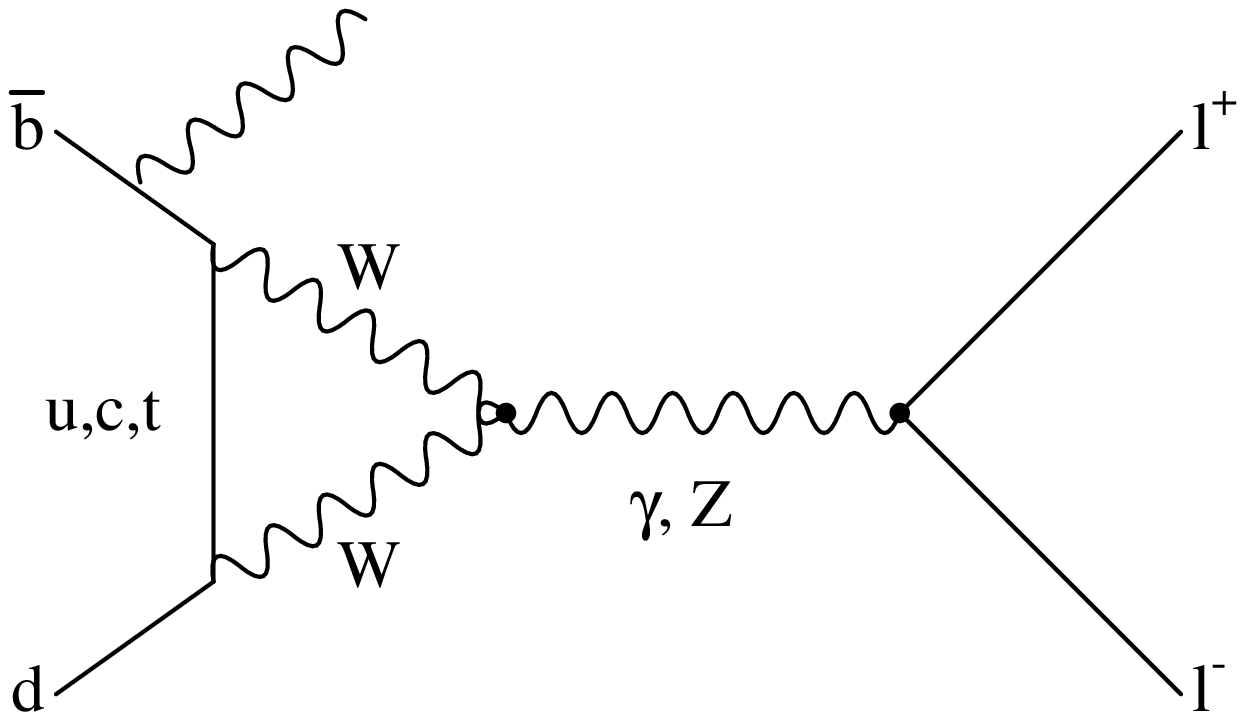, width=0.3\textwidth}}
\subfigure[]{\epsfig{file=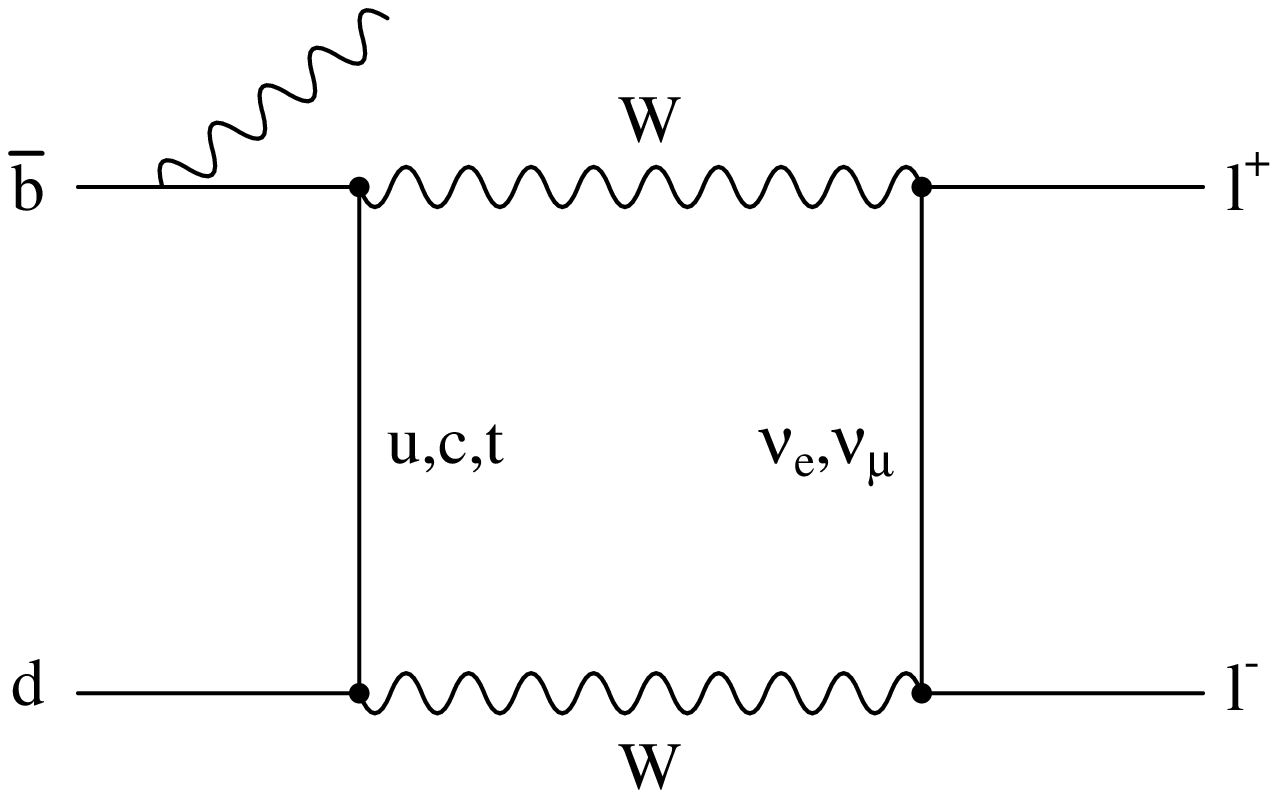, width=0.3\textwidth}}
\caption[Feynman Diagrams]{Feynman diagrams for the main contributions to \bllg decay. The signal photon can also 
be radiated from the initial $d$ quark line.}
\label{fig:feynman-diagram}
\end{center}
\end{figure}

The most important contribution to \bllg arises from radiative corrections to the pure leptonic processes 
$B^0\rightarrow\ell^+\ell^-$ which suffers from the helicity suppression. The short distance contributions to 
$B^0\rightarrow\ell^+\ell^-$ come from the box, Z-boson and photon-mediated diagrams. If a photon is emitted from the 
final charged lepton lines, the amplitude is proportional to the lepton mass $m_\ell$ and is helicity suppressed. When a photon 
is attached to any charged internal line, the contributions are also significantly suppressed by a factor of $m_b^2/m_W^2$. Therefore, the 
main contribution is when a photon is radiated from the initial quark lines as shown in \hjfig~\ref{fig:feynman-diagram}.

The expected branching fractions are summarized in \hjtab~\ref{tab:BR} for both radiative and non-radiative dilepton \B 
decays\cite{ref:theory-BRs}. This note presents the first search for \bllg.

\begin{table}[!htb]
 \caption[Branching Fraction Predictions] {The predicted Standard Model branching fractions.}
 \begin{center}
 \begin{tabular}{|c|r|}
 \hline Channel & Branching Fraction \\ \hline\hline
 \beeg & $1.5 - 4\times 10^{-10}$\\
 \bmmg & $1.2 - 3\times 10^{-10}$\\ \hline
 $\B^0\rightarrow e^+e^-$ & $2.1 \times 10^{-15}$ \\
 $\B^0\rightarrow \mu^+\mu^-$ & $9\times 10^{-11}$ \\ \hline
 \end{tabular}
 \end{center}
\label{tab:BR}
\end{table}

\section{THE \babar\ DETECTOR AND DATASET}
\label{sec:babar}
The data used in this analysis were collected with the \babar\ detector
at the \pep2\ storage ring at the Stanford Linear Accelerator Center and correspond to an integrated luminosity 
of \lumi\invfb accumulated at the \FourS resonance, which is equivalent to \nBB million \BB events,
and 27\invfb accumulated at a center-of-mass (\hjCM) energy about 40\mev below the \FourS resonance.

The \babar\ detector is described elsewhere~\cite{ref:babar}. The 1.5 T superconducting solenoidal magnet 
contains a charged particle tracking system with a silicon vertex tracker (SVT) followed by a drift chamber (DCH), 
a ring imaging Cherenkov detector (DIRC) 
dedicated to charged-particle identification, and an electromagnetic CsI(Tl) calorimeter (EMC). The 
segmented flux return (IFR) is instrumented with resistive plate chambers. About 1/3 of these chambers 
have been replaced with limited streamer tubes which provide higher muon identification 
efficiency. This change affects the most recent 77 \invfb of data.

A full \babar detector Monte Carlo (MC) simulation based on \texttt{GEANT4}\cite{ref:geant4} is used to evaluate 
signal efficiencies and to identify and study background sources.

\section{ANALYSIS METHOD}
\label{sec:Analysis}

We reconstruct \Bz candidates with two oppositely-charged leptons and a photon.  The leptons are required to originate 
from a common vertex, and the \Bz candidate is required to be consistent with coming from the beam interaction point. Since the 
signal events contain two neutral \B mesons and no additional particles, the total energy of each \B meson in the \hjCM frame must 
be equal to half of the total beam energy in the \hjCM frame. We define
\begin{align}
\label{eq:mes}
\mes&=\sqrt{(E_{\text{beam}}^*)^2-(\sum_i {\bf p}_i^*)^2}\\
\dE&=\sum_i\sqrt{m_i^2+({\bf p}_i^*)^2}-E_{\text{beam}}^*,
\end{align}
where $E_{\text{beam}}^*$ is the ($e^+$ or $e^-$) beam energy in the \hjCM frame, ${\bf p_i^*}$ and $m_i$ are the momentum 
in the \hjCM frame and the mass of the daughter particle $i$ ($i=\ell^+,\ell^-,\gamma$), respectively. In \hjeq~\ref{eq:mes}, 
$E_{\text{beam}}^*$ is used instead of the \B meson energy in the \hjCM frame because $E_{\text{beam}}^*$ 
is more precisely known. For correctly reconstructed \Bz 
mesons, \mes has a maximum at the nominal \Bz mass with a resolution of about 3\mevcc and \dE is near zero with a resolution 
of about 30\mev. 

The \bllg candidates are selected in the $-0.5\leq \dE\leq 0.5\gev$ and $5.2\leq\mes\leq 5.3\gevcc$ range. 
The size of the \SigBox is chosen to be approximately $\pm 3\sigma$ of the \dE and 
\mes distributions: $|\dE|\leq 0.123\gev$ and $5.27\leq \mes\leq 5.288\gevcc$. The resolutions in \dE and \mes are obtained 
from fits to the signal MC distributions assuming Crystal Ball function\cite{ref:CB} shapes. A larger region which contains the signal 
box was blinded during the development of this analysis: $|\dE|\leq 0.164\gev$ and $5.267\leq \mes\leq 5.29\gevcc$.
The remaining region, the \SideBox, is retained for background studies and data/MC comparison. 

To minimize the number of mis-identified particles, the leptons are required to satisfy 
stringent electron and muon identification criteria. Electrons are identified using a likelihood method with EMC and DIRC 
information. The electron identification efficiency is about 93\% and a pion fake rate is less than 0.1\%. 
Muons are identified using a neural network method with IFR information. The muon identification efficiency 
is about 70\% and a pion fake rate is less than 3\%. For photons, we require the shower lateral moment\cite{ref:lateral} 
to be less than $0.53(0.51)$ for the electron(muon) mode.

Leptons and photons are required to respect strict acceptance criteria. We have photon quality requirements on the number 
of crystals($\geq 10$) and photon energy($\geq 0.3\gev$) for the electron mode only. These requirements help reject beam 
background, the initial state radiations, and higher order QED backgrounds which are not modeled in MC for the electron mode. 

We apply vetoes on any leptons which may come from \jpsi or \psitwos decays. These vetoes are 
chosen in the 2-dimensional plane of \dE and the invariant mass of the two leptons, \mDL. The \jpsi veto region is the union 
of the following three regions in the \dE-\mDL plane for electron(muon) mode:
\begin{itemize}
\item $2.90 (3.00) < \mDL < 3.20 \gevcc$,
\item for $\mDL>3.20\gevcc$ region, a band in the $\dE-\mDL$ plane defined by \\
$1.11 c^2\times\mDL-3.58 (3.53)\gev<\dE<1.11 c^2\times\mDL-3.25 (3.31)\gev$,
\item for $\mDL<2.90(3.00)\gevcc$ region, a triangle in the $\dE-\mDL$ plane defined by\\
$\dE<1.11 c^2\times\mDL-3.25(3.31)\gev$.
\end{itemize}
Photon candidates whose invariant mass with other photons
in the event are in the range of $0.115<m_{\gamma\gamma}<0.155$\gevcc are vetoed to remove photons from \piz decays. We 
studied the effect of an $\eta$ veto and found it to be negligible. 

\begin{figure}[!htb]
\begin{center}
\subfigure[]{\epsfig{file=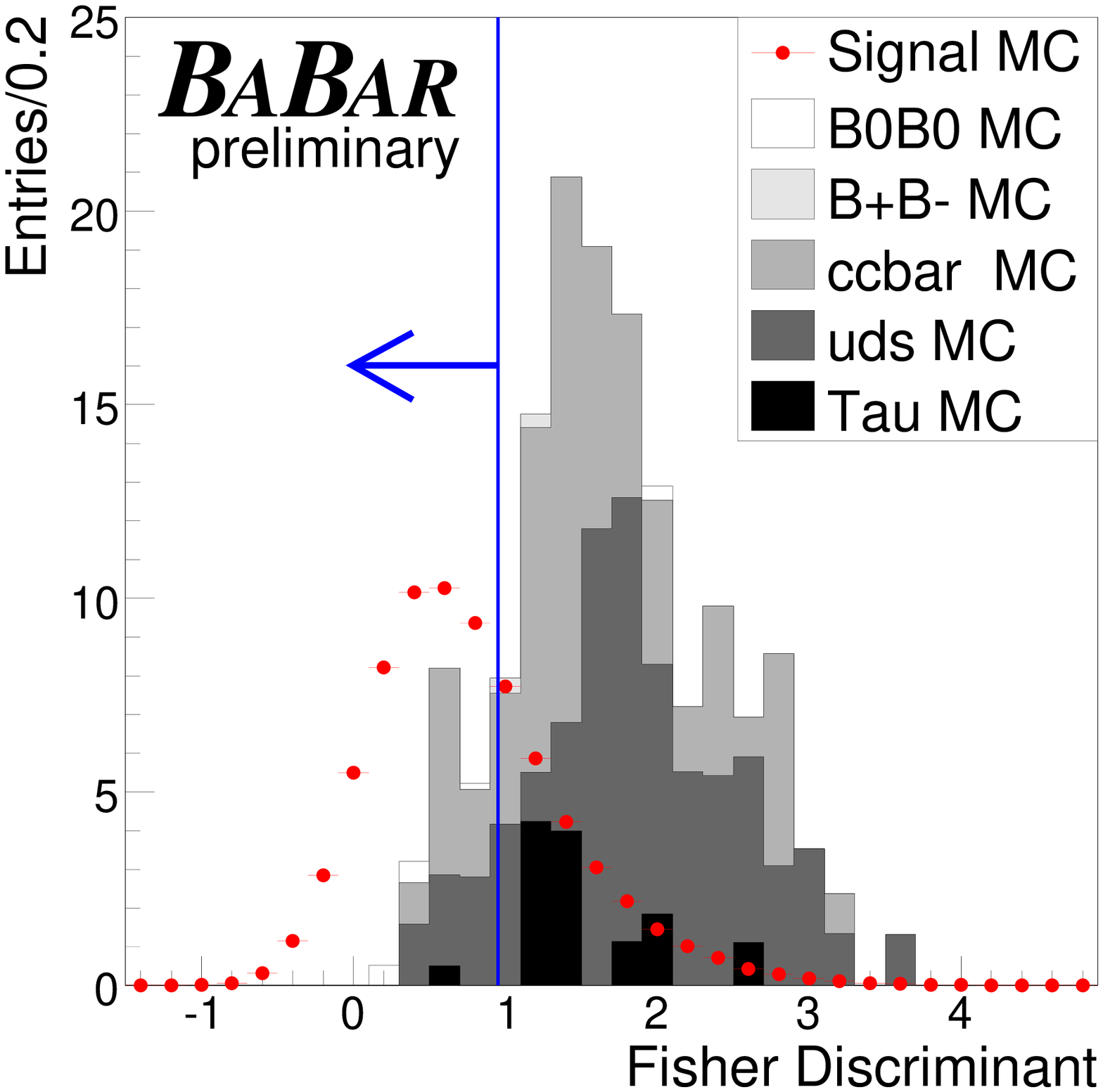,  height=6cm, angle=0}}\hspace{0.1in}
\subfigure[]{\epsfig{file=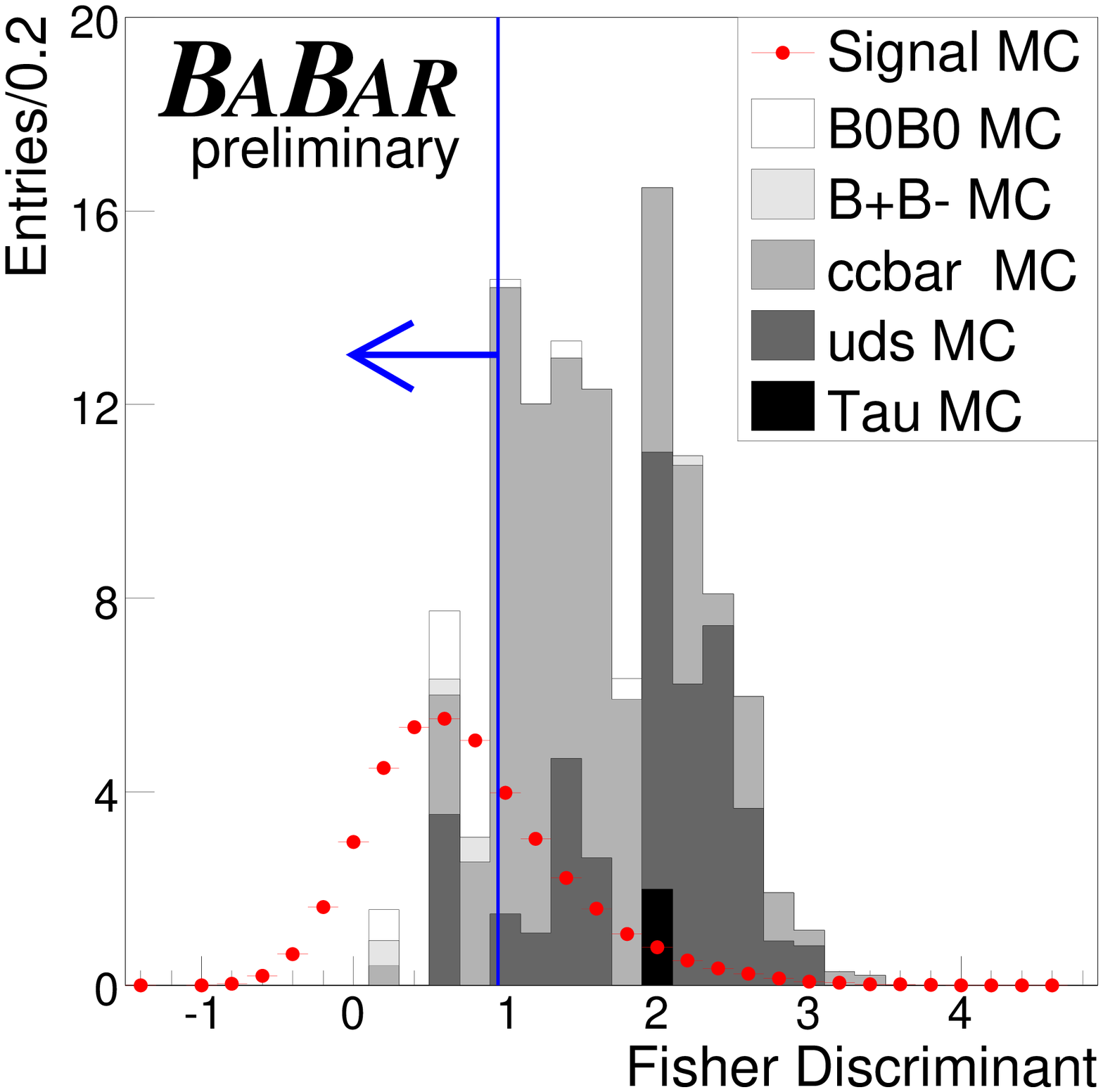, height=6cm, angle=0}}
\caption[]{The distribution of the Fisher discriminant for selected events in the signal box for (a) \beeg and 
(b) \bmmg sample. The points show the predicted distributions for signal events, the histograms show the various 
background contributions. The \jpsi, \psitwos and \piz vetoes are applied. The 
signal distribution is normalized to the branching fraction of $1\times 10^{-6}$. 
}
\label{fig:Fisher}
\end{center}
\end{figure}

We require that the invariant mass of dilepton system is between 0.3 and $4.9(4.7)\gevcc$ for the electron(muon) mode to reject 
non-\BB background. Further suppression of background from non-\BB background is provided by a series of topological requirements. 
We require $|\cos\theta_T|\leq 0.8$, where $\theta_T$ is the angle in the \hjCM frame between the thrust axis of 
the particles that form the reconstructed \Bz candidate and the thrust axis of the remaining tracks and neutral clusters 
in the event. We have a requirement on the ratio of the second to zeroth Fox-Wolfram 
moment\cite{ref:Fox-Wolfram} of $R_2\leq 0.31$. We define a Fisher discriminant based on the following variables: 
the angle between the \Bz direction and the 
beam axis, the angle between the thrust axis of the \Bz candidate and the beam axis, and the summed momentum of 
the rest of the event tracks and neutrals in nine non overlapping volumes delimited by cones centered around 
the thrust axis of the \Bz candidate with half-angles in $10^\circ$ steps. The distributions 
of the Fisher discriminants are shown in \hjfig~\ref{fig:Fisher}. The arrows show the allowed regions. If multiple \Bz candidates 
pass all the selection requirements in an event, only the one with \dE closest to zero is retained. The average number of \Bz 
candidates per event is 1 (1.05) for the electron(muon) mode.

To assess potential background contributions peaking like the signal in \dE and \mes, we examined 32 exclusive hadronic and 
semileptonic decay modes, and found no significant contribution.

The requirement levels are optimized by minimizing the expected upper limit branching fraction at 90\% confidence level (C.L.). 
The upper limit is calculated using the Feldman-Cousins method\cite{ref:FeldmanCousins}. The normalizations and shapes of the 
data and MC distributions are compared at various requirement levels in the sideband area. They 
show reasonable agreement at all levels.

\begin{figure}[!htb]
\begin{center}
\subfigure[]{\epsfig{file=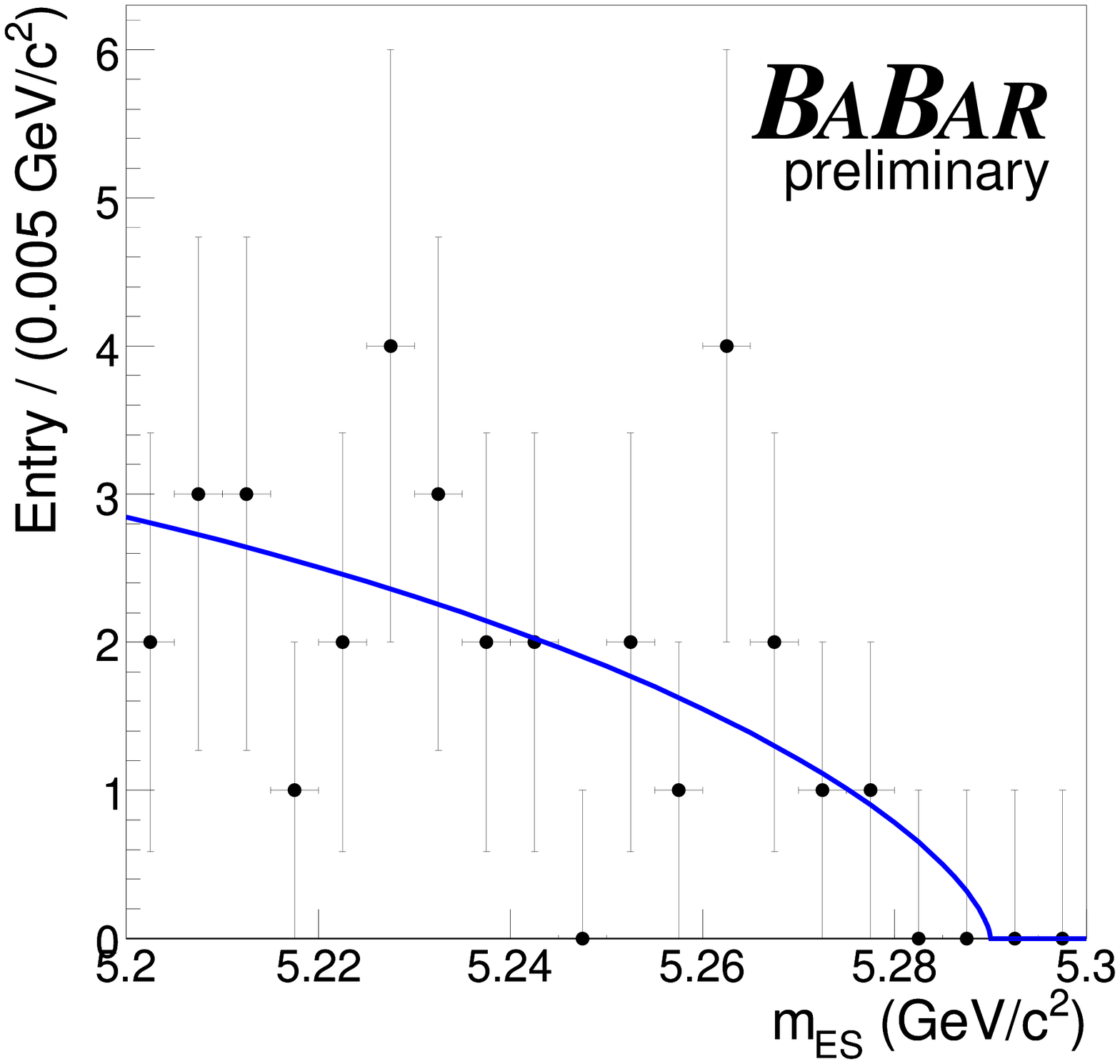, height=5.8cm}} \hspace{0.1in}
\subfigure[]{\epsfig{file=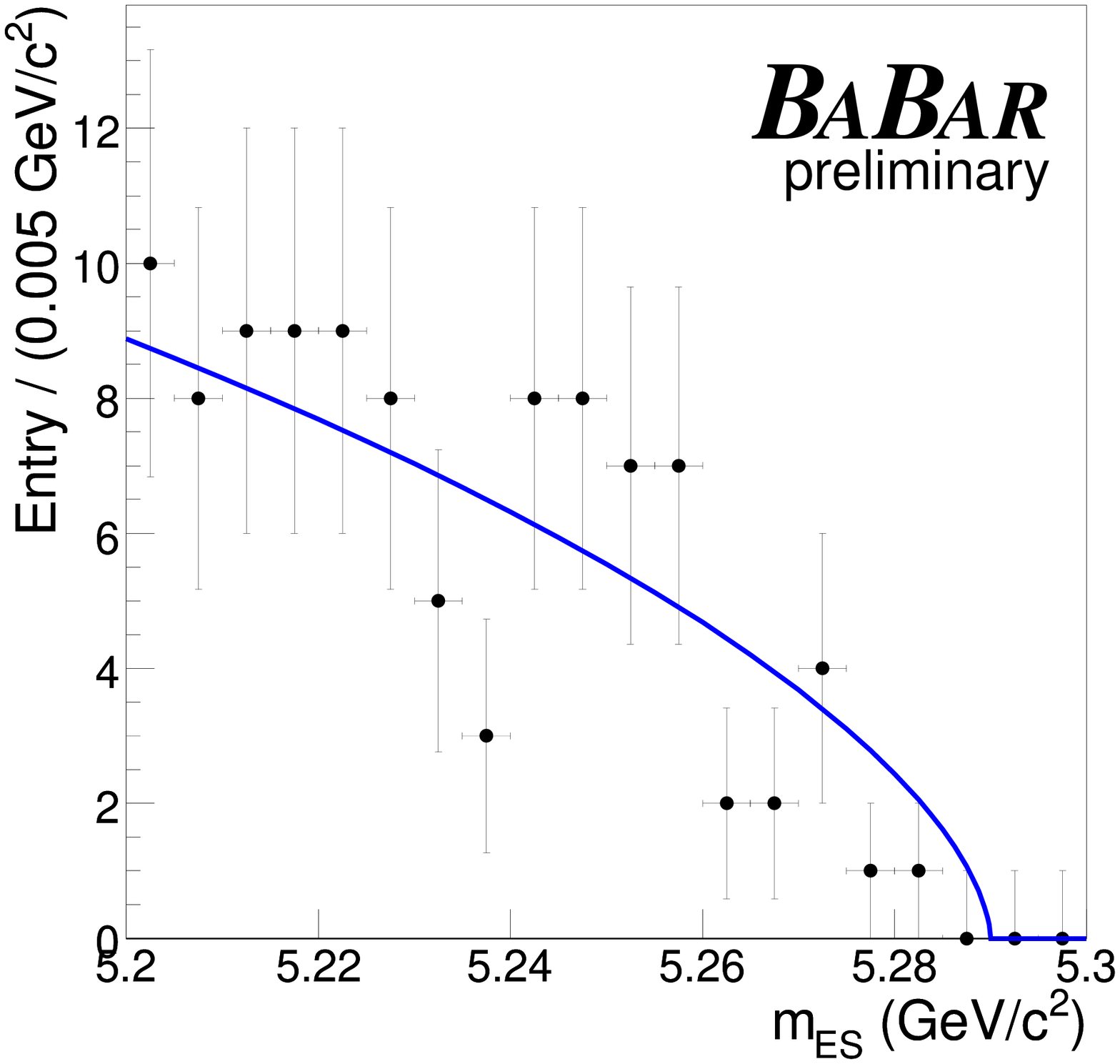, height=5.8cm}}
\caption[]{An unbinned maximum likelihood fit on \mes distribution in the combined upper and lower sideband boxes, 
using an Argus function for (a) \beeg and (b) \bmmg sample. The points represent data and the solid lines show the fit function.}
\label{fig:Fit}
\end{center}
\end{figure}

To estimate the background level in the signal box, three different sideband boxes are used, as indicated in 
\hjtab~\ref{tab:boxes}. We perform an unbinned maximum likelihood fit on the \mes distribution in the combined upper 
and lower sideband boxes with an Argus function\cite{ref:argus}, as shown in \hjfig~\ref{fig:Fit}. 
The end-point of the Argus function is fixed at 5.29\gevcc. We 
use this parameterization to extrapolate the background level found in the middle sideband (12 events for \beeg sample and 
13 events for \bmmg sample) into the signal box. The error on the expected number of background in the signal box from the 
fit is estimated by varying the Argus parameters by $\pm 1 \sigma$. The background has been estimated with alternative 
methods, using sidebands in \dE, and repeating the fit after relaxing the selection requirements. The estimates obtained 
are in general compatible within errors. The fit to the \mes sidebands was chosen as the preferred method because it yields 
the smallest statistical uncertainties.

\begin{table}[!htb]
 \caption[] {Definitions of the different sideband boxes used.}
 \begin{center}
 \begin{tabular}{|c|c|c|}
 \hline Sideband Box & span in \dE (\gev) & span in \mes (\gevcc) \\ \hline\hline
 Upper Sideband & (0.164, 0.5)  &  (5.2, 5.3) \\
 Lower Sideband & (-0.5, -0.164) & (5.2, 5.3)\\ 
 Middle Sideband & (-0.123, 0.123) & (5.2, 5.26)\\ \hline
 \end{tabular}
 \end{center}
\label{tab:boxes}
\end{table}

\section{SYSTEMATIC STUDIES}
\label{sec:Systematics}
Systematic uncertainties are evaluated using independent data control samples. The dominant systematic uncertainty 
is related to the reconstruction of the signal photon energy which is determined using  $e^+e^-\rightarrow\mu^+\mu^-\gamma$ decays. 
The uncertainty is 1.6\% for both modes. The systematic uncertainty from the particle identification 
has been determined using an independent control sample of \jpsi decays. It is 0.7\% and 1.3\% for the electron and muon mode 
respectively. The uncertainty on the number of \BB events is 1.1\%\cite{ref:B-counting}. The tracking 
efficiency is determined from $e^+e^-\rightarrow\tau^+\tau^-$ interactions, 
with one tau decaying leptonically and the other to three charged hadrons. The uncertainty is 0.94\% for both 
electron and muon modes. All contributions to the systematics are added in quadrature 
to give a total relative systematic uncertainty of 2.3\% for the electron mode and 2.5\% for the muon mode on 
the branching fraction.

\section{RESULTS}
\label{sec:Physics}
\begin{figure}[!htb]
\begin{center}
\subfigure[]{\epsfig{file=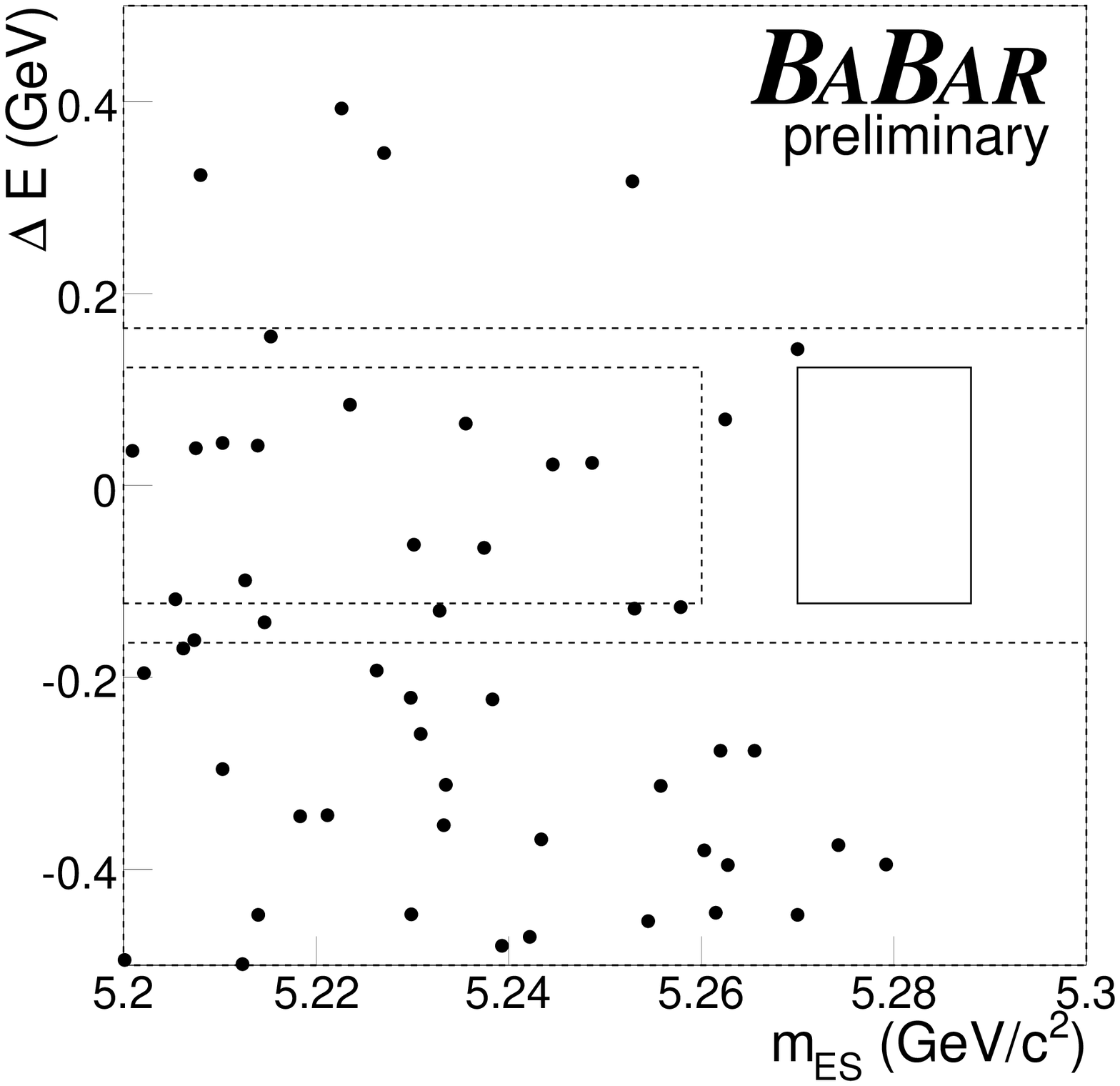, height=6cm}} \hspace{0.1in}
\subfigure[]{\epsfig{file=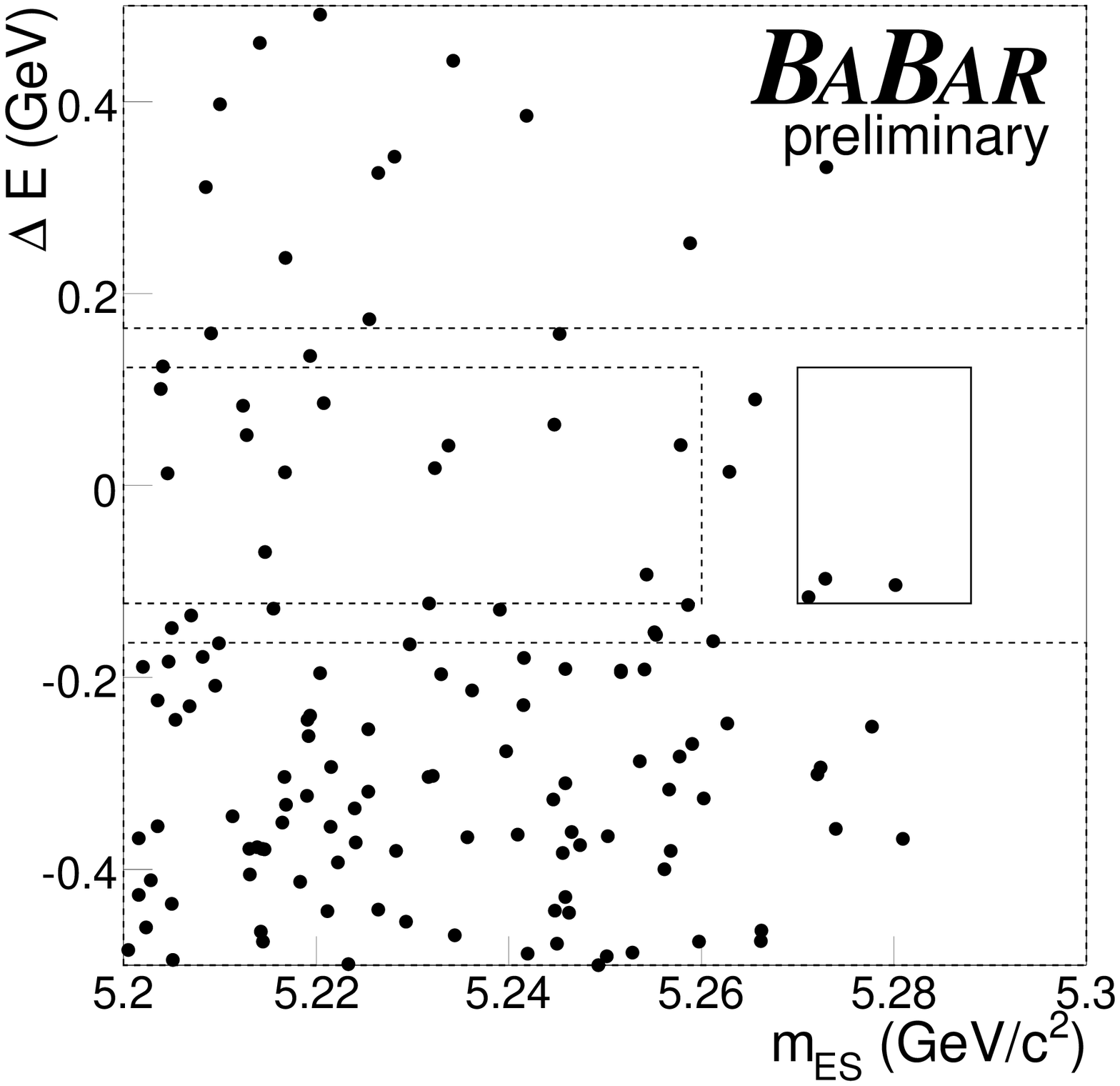, height=6cm}}
\caption[]{Distribution of events in \mes and \dE for (a) \beeg and (b) \bmmg sample. 
The solid box represents the signal box. The three dashed-lined area show the upper, middle, and lower sideband boxes
from top to bottom. }
\label{fig:result-data}
\end{center}
\end{figure}

As shown in \hjfig~\ref{fig:result-data} and  \hjtab~\ref{tab:result}, 
zero and three events were found in the electron and muon modes, respectively.  The number of events 
found in the signal box is compatible with the expected background for both modes. 

An upper limit on the branching fraction is computed using
\begin{equation}
\BR_{UL}(\bllg)=\frac{N_{UL}}{N_{\Bz}\cdot\effsig}
\end{equation}
where $N_{UL}$ is the 90\% C.L. upper limit for the signal yield as determined by the method described in \cite{ref:Barlow} 
with both statistical and systematic errors, $N_{\Bz}$  is the number of the neutral \B mesons and \effsig is the signal 
reconstruction efficiency from the signal MC sample. $N_{\Bz}$ is equal to the number of \BB events, as we are assuming 
$\BR(\FourS\rightarrow\BzBzb)=\BR(\FourS\rightarrow\BpBm)$. The obtained 90\% C.L. branching fraction upper limits 
are $\BR(\beeg)<\ResultE\times 10^{-7} $and $\BR(\bmmg)<\ResultM \times 10^{-7}$.

\begin{table}[!htb]
 \caption[] {Summary of the results where \nobs and \nbgexp are the observed and expected number of background events in the 
signal box, \effsig is the efficiency, $N_{UL}$ is the 90\% C.L. upper limit for the signal yield, and $\BR_{UL}(\bllg)$ 
is the upper limit on the branching fraction at the 90\% C.L. 
Systematic uncertainties on \nbgexp and \effsig are given. }
 \begin{center}
 \begin{tabular}{|c|c|c|c|c|c|}  \hline
 Decay Mode & \nobs & \nbgexp & \effsig (\%) & $N_{UL}$ & $\BR_{UL}(\bllg)$\\  \hline\hline
 $e^+e^-$           & 0 & $1.28 \pm 0.80$ & $6.07 \pm 0.14$ & 1.33 & $\ResultE \times 10^{-7}$\\
 $\mu^+\mu^-$ & 3 & $1.40 \pm 0.42$ &  $4.93 \pm 0.12$ & 5.30 & $\ResultM \times 10^{-7}$\\ \hline
 \end{tabular}
 \end{center}
\label{tab:result}
\end{table}

\section{CONCLUSION}
\label{sec:Summary}
A search for the \bllg ($\ell=e$ or $\mu$) modes has been performed based on \nBB million \BB events. We 
obtain 90\% C.L. upper limits for the branching fraction of $\BR(\beeg)<\ResultE\times 10^{-7} $and 
$\BR(\bmmg)<\ResultM \times 10^{-7}$. These are the only limits currently available for these decay modes.

\section{ACKNOWLEDGMENTS}
\label{sec:Acknowledgments}

We are grateful for the 
extraordinary contributions of our \pep2\ colleagues in
achieving the excellent luminosity and machine conditions
that have made this work possible.
The success of this project also relies critically on the 
expertise and dedication of the computing organizations that 
support \babar.
The collaborating institutions wish to thank 
SLAC for its support and the kind hospitality extended to them. 
This work is supported by the
US Department of Energy
and National Science Foundation, the
Natural Sciences and Engineering Research Council (Canada),
Institute of High Energy Physics (China), the
Commissariat \`a l'Energie Atomique and
Institut National de Physique Nucl\'eaire et de Physique des Particules
(France), the
Bundesministerium f\"ur Bildung und Forschung and
Deutsche Forschungsgemeinschaft
(Germany), the
Istituto Nazionale di Fisica Nucleare (Italy),
the Foundation for Fundamental Research on Matter (The Netherlands),
the Research Council of Norway, the
Ministry of Science and Technology of the Russian Federation, and the
Particle Physics and Astronomy Research Council (United Kingdom). 
Individuals have received support from 
the Marie-Curie IEF program (European Union) and
the A. P. Sloan Foundation.

\end{document}